\documentclass[a4paper]{article}

\PassOptionsToPackage{numbers, compress}{natbib}

\usepackage{INTERSPEECH2020}
\usepackage{multirow}
\usepackage[utf8]{inputenc} 
\usepackage[T1]{fontenc}    
\usepackage{hyperref}       
\usepackage{url}            
\usepackage{booktabs}       
\usepackage{amsfonts}       
\usepackage{nicefrac}       
\usepackage{microtype}      

\usepackage{graphicx}
\usepackage{subfigure}
\usepackage{booktabs} 
\usepackage{multirow}
\usepackage{siunitx}
\usepackage{float}

\usepackage{amsmath,amsthm,amssymb,bbm,stmaryrd,bm}
\usepackage{xargs}                      
\usepackage[pdftex,dvipsnames]{xcolor}  
\usepackage[colorinlistoftodos,prependcaption,textsize=tiny]{todonotes}
\newcommandx{\unsure}[2][1=]{\todo[linecolor=red,backgroundcolor=red!25,bordercolor=red,#1]{#2}}
\newcommandx{\change}[2][1=]{\todo[linecolor=blue,backgroundcolor=blue!25,bordercolor=blue,#1]{#2}}
\newcommandx{\info}[2][1=]{\todo[linecolor=OliveGreen,backgroundcolor=OliveGreen!25,bordercolor=OliveGreen,#1]{#2}}
\newcommandx{\improvement}[2][1=]{\todo[linecolor=Plum,backgroundcolor=Plum!25,bordercolor=Plum,#1]{#2}}
\newcommandx{\thiswillnotshow}[2][1=]{\todo[disable,#1]{#2}}
\newcommandx{\fix}[2]{{\color{red} #1}\todo{#2}}

\def\x{{\mathbf x}}
\def\X{{\mathbf X}}

\def\cf{{\mathbf c}}
\def\m{{\mathbf m}}

\def\CF{{\mathbf C}}
\def\M{{\mathbf M}}
\def\H{{\mathbf H}}
\renewcommand{\S}{\mathcal{S}}
\newcommand{\IS}{\mathcal{S}^{\dagger}}

\usepackage{array}
\usepackage{booktabs} 
\usepackage{bigstrut}
\makeatletter
\newlength\mylena
\newlength\mylenb
\newcommand\mystrut[1][2]{%
    \setlength\mylena{#1\ht\@arstrutbox}%
    \setlength\mylenb{#1\dp\@arstrutbox}%
    \rule[\mylenb]{0pt}{\mylena}}
\makeatother

    \newcommand{\Lc}{\mathcal{L}}

\title{Hide and Speak: Towards Deep Neural Networks for Speech Steganography}

\name{Felix Kreuk$^{1}$, Yossi Adi$^{2}$, Bhiksha Raj$^{3}$, Rita Singh$^{3}$, Joseph Keshet$^{1}$}
\address{$^1$Bar-Ilan University~~~~~~~~~$^2$Facebook AI Research~~~~~~$^3$Carnegie Mellon University}
\email{felix.kreuk@gmail.com}

\begin{document}

\maketitle

\begin{abstract}
Steganography is the science of hiding a secret message within an ordinary public message, which is referred to as Carrier. Traditionally, digital signal processing techniques, such as least significant bit encoding, were used for hiding messages. In this paper, we explore the use of deep neural networks as steganographic functions for speech data. We showed that steganography models proposed for vision are less suitable for speech, and propose a new model that includes the short-time Fourier transform and inverse-short-time Fourier transform as differentiable layers within the network, thus imposing a vital constraint on the network outputs.
We empirically demonstrated the effectiveness of the proposed method comparing to deep learning based on several speech datasets and analyzed the results quantitatively and qualitatively. Moreover, we showed that the proposed approach could be applied to conceal multiple messages in a single carrier using multiple decoders or a single conditional decoder. Lastly, we evaluated our model under different channel distortions. Qualitative experiments suggest that modifications to the carrier are unnoticeable by human listeners and that the decoded messages are highly intelligible.
\end{abstract}

\section{Introduction}

Steganography (``steganos'' -- concealed or covered, ``graphein'' -- writing) is the science of concealing messages inside other messages.  It is generally used to convey concealed ``secret'' messages to recipients who are aware of their presence, while keeping even their existence hidden from other unaware parties who only see the ``public'' or ``carrier'' message.  

Recently, \cite{baluja2017hiding, zhu2018hidden} proposed to use deep neural networks as a steganographic function for hiding an image inside another image. Unlike traditional steganography methods~\cite{morkel2005overview, kessler2004overview}, in this line of work, the network {\em learns} to conceal a hidden message inside the carrier without manually specifying a particular redundancy to exploit. 

Although these studies presented impressive results on image data, the applicability of such models for speech data was not explored. As opposed to working with raw images in the domain of vision processing, the common approach when learning from speech data is to work at the frequency domain, and specifically, using the short time Fourier transform (STFT) to capture the spectral changes over time. The STFT output is a complex matrix composed of the Fourier transform of different time frames. The common practice is to use the absolute values (magintudes) of the STFT measurements, and to maintain a substantial overlap between adjacent frames~\cite{lim1979enhancement}. Consequently, the original signal cannot be losslessly recovered from STFT. Moreover, as only the magnitude is considered, the phase needs to be recovered. This process complicates the restoration of the time domain signal even further.

In this study, we show that steganography models proposed for vision are less suitable for speech. We build on the work by \cite{baluja2017hiding, zhu2018hidden} and propose a new model that includes the STFT and inverse-STFT as differentiable layers within the network, thus imposing a vital constraint on the network outputs. 

Although one can simply hide written text inside audio files and convey the same lexical content, concealing audio inside audio preserves additional features. For instance, the secret message may convey the speaker identity, the sentiment of the speaker, prosody, etc. These features can be used for later identification and authentication of the message.

Similarly to~\cite{baluja2017hiding, zhu2018hidden}, the proposed model is comprised of three parts. The first learns to encode a hidden message inside the carrier. The second component are differential STFT and inverse-STFT layers that simulate transformations between frequency and time domains. Lastly, the third component learns to decode a hidden message from a generated carrier. Additionally, we demonstrated for the first time, that the above scheme now permits us to hide {\em multiple} secret messages into a \emph{single} carrier, each potentially with a different intended recipient who is the only person who can recover it.

Further analysis shows that the addition of STFT layers yields a method which is robust to various channel distortions and compression methods, such as MP3 encoding, Additive White Gaussian Noise, sample rate reduction, etc. 
Qualitative experiments suggest that modifications to the carrier are unnoticeable by human listeners and that the decoded messages are highly intelligible and preserve other semantic content, such as speaker identity.

\paragraph*{Our contribution:}
\begin{itemize}
\item We empirically show that steganographic vision-oriented models are less suitable for the audio domain.
\item We augment vision-oriented models with differentiable STFT/Inverse-STFT layers during training to care for noise introduced when converting signals from frequency to time domain and back.
\item We embed multiple speech messages in a single speech carrier.
\item We provide extensive empirical and subjective analysis of the reconstructed signals and show that the produced carriers are indistinguishable from the original carriers, while keeping the decoded messages highly intelligible. 
\end{itemize}

\noindent The paper is organized as follows, Section~\ref{sec:prob_set} formulates all the notations we use throughout the paper. In Section~\ref{sec:model} we describe the proposed model. Section~\ref{sec:results} and Section~\ref{sec:analysis} present the results together with objective and subjective analysis. Section~\ref{sec:rel_work} summarizes the related work. We conclude the paper in Section~\ref{sec:dis} with a discussion and future work.


\section{Notation and representation}
\label{sec:prob_set}

In this section, we rigorously set the notation we use throughout the paper. 

\paragraph*{Steganography notations.} Recall, in steganography the goal is to conceal a hidden message within a carrier segment. Specifically, the steganography system is a function that gets as input a carrier utterance, denoted by $\cf$, and a hidden message, denoted by $\m$. The outputs of the system are the \emph{embedded carrier} $\hat{\cf}$, and consequently the \emph{recovered message}, $\hat{\m}$, such that the following constraints are satisfied: (i) both $\hat{\cf}$ and $\hat{\m}$ should be perceptually similar to $\cf$ and $\m$, respectively, by a human evaluator; (ii) the message $\hat{\m}$ should be recoverable from the carrier $\hat{\cf}$ and should be intelligible; and lastly (iii) a human evaluator should not be able to detect the presence of a hidden message embedded in $\hat{\cf}$. 

\paragraph*{Audio notations.} Let $\x = (x[0], x[1], \ldots,  x[N-1])$ be a speech signal that is composed of $N$ samples. The spectral content of the signal changes over time, therefore it is often represented by the short-time Fourier transform, commonly known as the \emph{spectrogram}, rather than by the Fourier transform. 

The STFT, $\mathcal{X}$, is a matrix of complex numbers, its columns are the Fourier transform of a given time frame and its rows are frame indices. In speech processing we are often interested in the  absolute value of the STFT, or the magnitude, which is denoted by $\X=|\mathcal{X}|$. Similarly we denote the phase of the STFT by $\angle \X = \arctan\left(\operatorname{Im}(\mathcal{X})/\operatorname{Re}(\mathcal{X})\right)$. Furthermore, we denote by $\S$ the operator that gets as input a real signal and outputs the magnitude matrix of its STFT, $\X=\S(\x)$, and denote by $\IS$ the operator that gets as input the magnitude and phase matrices of the STFT, and returns a recovered version of the speech waveform, $\x=\IS(\X,\angle \X)$. Here $\IS$ is computed by taking the inverse Fourier transform of each column of $\X$, and then reconstructing the waveform by combining the outputs by the overlap-and-add method. Note that this reconstruction is imperfect, since there is a substantial overlap between adjacent windows when using STFT in speech processing, hence part of the signal at each window is lost \cite{jaganathan2016stft}. 
\section{Model}
\label{sec:model}

\begin{figure}
\centering
\includegraphics[width=\linewidth]{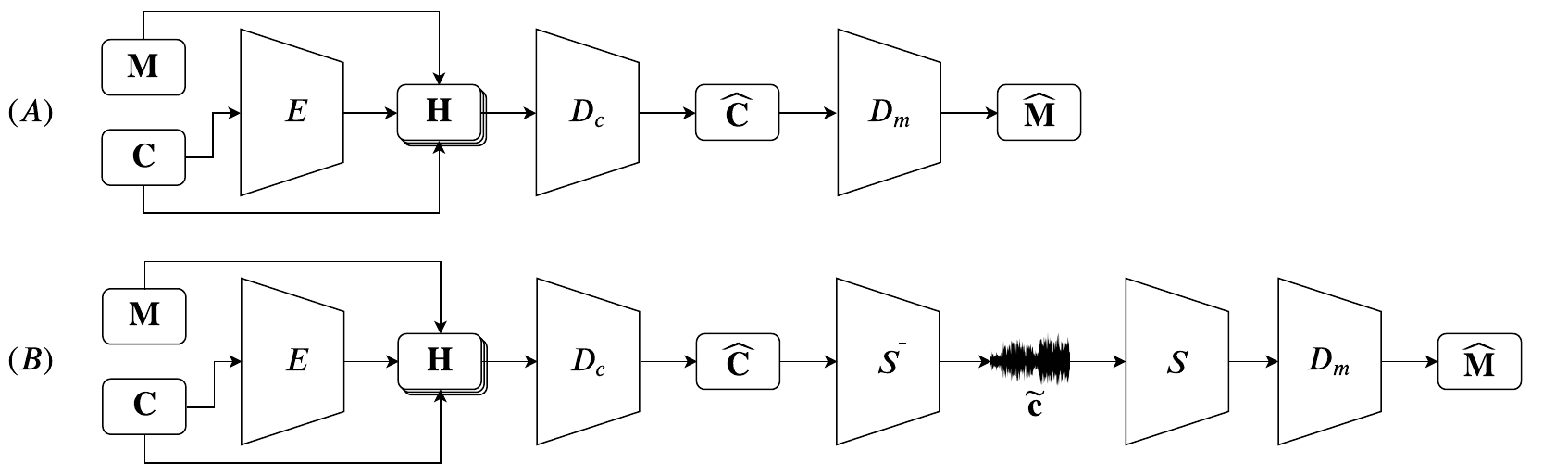}
\caption{Model overview: the encoder $E$ gets as input the carrier $\CF$, its output is then concatenated with $\CF$ and $\M$ to create $\H$. Then, the carrier decoder $D_c$ generates the new embedded carrier, from which the message decoder $D_m$ decodes the message $\hat{\M}$; Sub-figure (1A) depicts the baseline model (1B) depicts our proposed model.}
\label{fig:model}
\end{figure}

Similarly to the models proposed in \cite{baluja2017hiding,zhu2018hidden}, our architecture is composed of the following components: (i) \textit{Encoder Network} denoted $E$; (ii) \textit{Carrier Decoder Network} denoted $D_c$; and (iii) \textit{Message Decoder Network} denoted $D_m$. The model is schematically depicted in Figure~\ref{fig:model}A. The Encoder Network $E$, gets as input a carrier $\CF$, and outputs a latent representation of the carrier, $E(\CF)$. Then, we compose a joint representation of the encoded carrier $E(\CF)$, message $\M$, and original carrier $\CF$ by concatenating all three along the convolutional channel axis, $\H = [E(\CF); \CF; \M]$ as proposed in \cite{zhu2018hidden}, where we denote the concatenation operator by ;. 

The Carrier Decoder Network, $D_c$, gets as input the aforementioned representation and outputs $\hat{\CF}$, the carrier embedded with the hidden message. Lastly, the Message Decoder Network $D_m$, gets as input $\hat{\CF}$ and outputs $\hat{\M}$, the reconstructed hidden message. Each of the above components is a neural network, where the parameters are found by minimizing the absolute error between the carrier and the embedded carrier and between the original message and the reconstructed message. 

At this point our architecture diverges from the one proposed in~\cite{baluja2017hiding} by the addition of a differentiable STFT layers. Recall, our goal is to transmit $\hat{\cf}$, which means we need to recover the time-domain waveform from the magnitude $\hat{\CF}$. Unfortunately, the recovery of $\hat{\cf}$ from the STFT magnitude only, is an ill-posed problem in general \cite{hofstetter1964construction, jaganathan2016stft}. Ideally, we would like to reconstruct $\hat{\cf}$ using $\IS(\hat{\CF}, \angle \hat{\CF})$. However, the phase $\angle \hat{\CF}$ is unknown, and therefore must be approximated. 

One way to overcome this phase recovery obstacle is to use the classical alternating projection algorithm of Griffin and Lim \cite{griffin1984signal}. 
Unfortunately, this method produces a carrier with noticeable artifacts. The message, however, can be recovered that way and is intelligible.

Another way to reconstruct the time-domain signal is to use the magnitude of the embedded carrier $\hat{\CF}$, and the phase of the original carrier, $\angle \CF$. In subjective tests we found that the restructured carrier, denoted as $\tilde{\cf}=\IS(\hat{\CF}, \angle \CF)$, sounds acoustically similar to the original carrier $\cf$. However, when recovering the hidden message we get unintelligible output. This is due to the fact that we used a mismatched phase. 

To mitigate that, we turn to a third solution, where we constrain the loss function by $\S$ and $\IS$. 
Formally, we minimize: 
\[
\Lc(\CF, \M) = \lambda_c\|\CF - \hat{\CF}\|_1  + \lambda_m\|\M - \tilde{\M}\|_1,\\ 
\]
where  $\hat{\CF} = D_c(\H), \tilde{\CF} = \S(\IS(\hat{\CF}, \angle \CF))$, and $\tilde{\M}=D_m(\tilde{\CF})$.
Practically, we added $\S$ and $\IS$ operators as  differentiable 1D-convolution layers as illustrated in Figure~\ref{fig:model}B. In words, we jointly optimize the model to generate $\hat{\CF}$ which will preserve the hidden message after $\S\circ\IS$ and will also resemble $\CF$.

The above approach can be naturally extended to conceal multiple messages. In that case, the model is provided with a single carrier $\CF$, and a set of $k$ messages, $\{\M_i\}_{i=1}^k$, where $k>1$. We explored two settings: (i) \emph{multiple message decoders}, in which we use $k$ different message decoders denoted by $D_{m,i}$ where $1 < i\le k$, one for each message; and (ii) \emph{a single conditional decoder}, in which we condition a single decoder $D_{m}$ with a set of codes $\{q_i\}_{i=1}^k$. Each code $q_i$ is represented as a one-hot vector of size $k$.
\section{Experimental results}
\label{sec:results}
We evaluated our approach on TIMIT \cite{garofolo1993darpa} and YOHO \cite{campbell1995testing} datasets using the standard train/val/test splits. We evaluated the proposed method on the aforementioned datasets to assess the model under various recording conditions. Each utterance was sampled at 16kHz and represented as its power spectrum by applying the STFT with $W=256$ FFT frequency bins and sliding window with a shift $L=128$. Training examples were generated by randomly selecting one utterance as carrier and $k$ other utterances as messages for $k \in \{1,3,5\}$. Thus, the matching of carrier and message is completely arbitrary and not fixed. Further, it may originate from different speakers. 

All models were trained using Adam for 80 epochs with an initial learning rate of $10^{-3}$ and a decaying factor of 10 every 20 epochs. We balanced between the carrier and message reconstruction losses using $\lambda_c=3$, $\lambda_m=1$. Each component in our model is implemented as a Gated Convolutional Neural Network as proposed by \cite{dauphin2017language}. Specifically, $E$ is composed of three blocks of gated convolutions, $D_c$ was composed of four blocks of gated convolutions, and $D_m$ was composed of six blocks of gated convolutions. Each block contained 64 kernels of size 3$\times$3. Sample waveforms of different models and experiments as well as the source code are available at \url{http://hyperurl.co/ab7c3g}.

We report results for the proposed approach together with~\cite{baluja2017hiding, zhu2018hidden}. Additionally, we included a naive baseline, denoted by \emph{Frequency Chop}. In which, we concatenated the lower half of frequencies of $\M$ above the lower half of frequencies of $\CF$, to form $\hat{\CF}$.
Message decoding was performed by extracting the upper half of frequencies from $\hat{\CF}$ and zero padding to the original size.

Results for concealing a single message are reported in Table~\ref{tab:single_msg_res}: the Absolute-Error (AE) and Signal-to-Noise-Ratio (SNR) for both carrier and message of all baselines and the proposed models on TIMIT and YOHO.

\begin{figure*}
\centering
\includegraphics[width=0.24\linewidth]{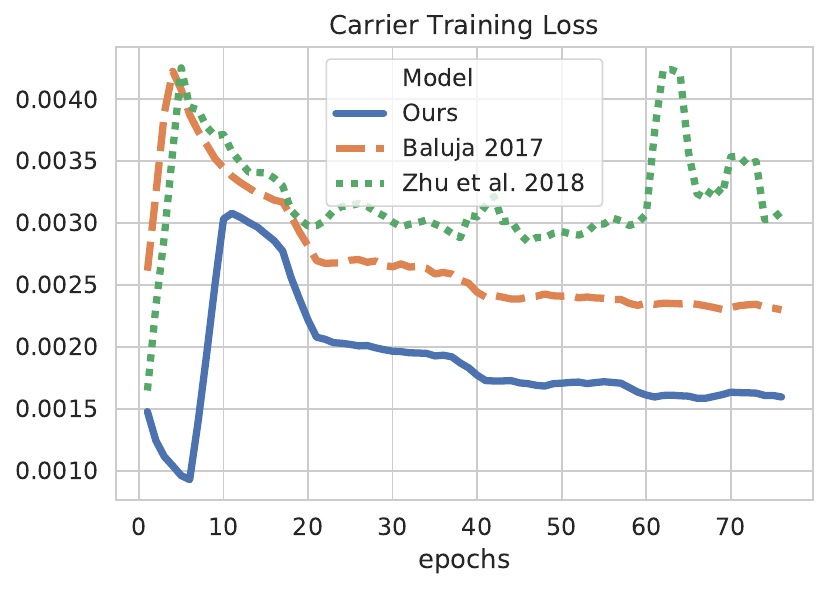} 
\includegraphics[width=0.24\linewidth]{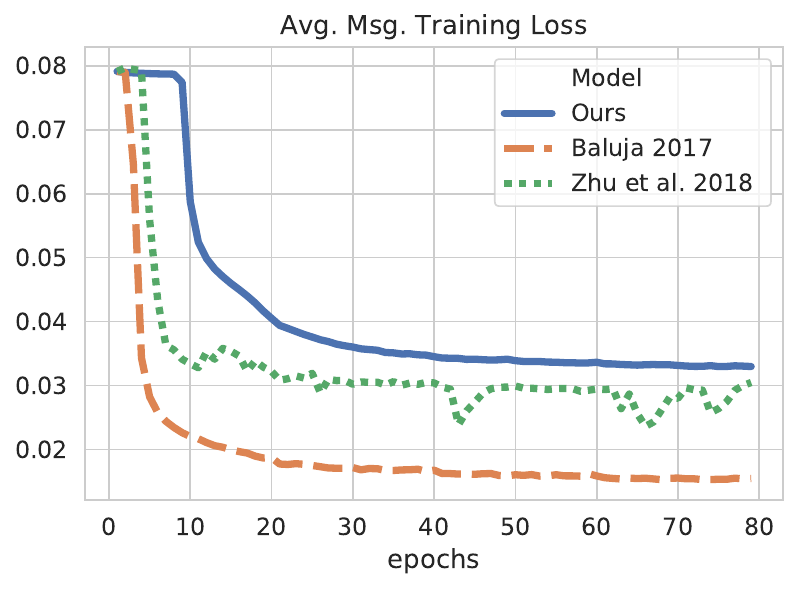}
\includegraphics[width=0.24\linewidth]{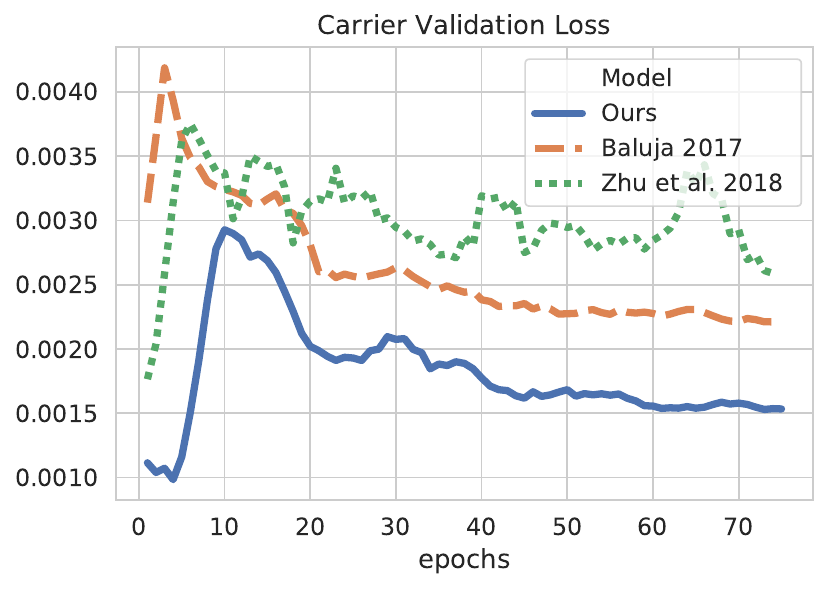} 
\includegraphics[width=0.24\linewidth]{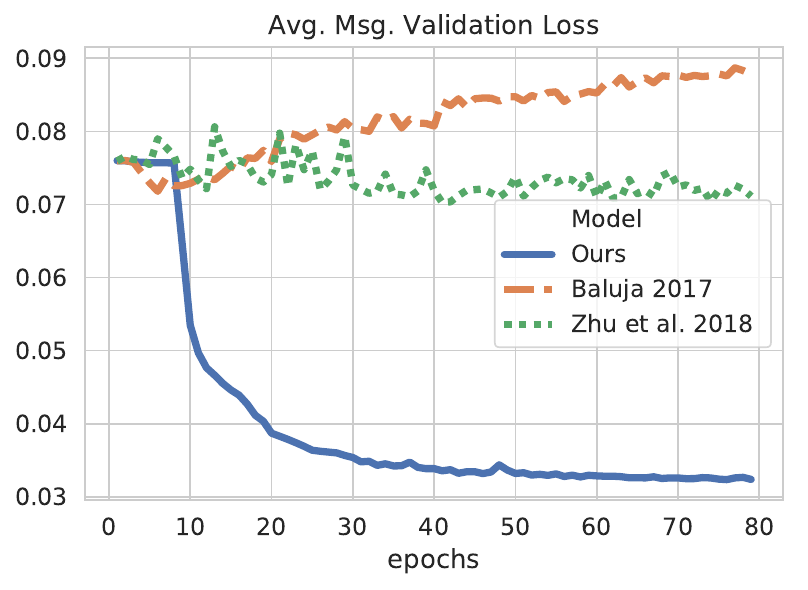}
\caption{
During validation, we simulate real world conditions by converting the carrier to time-domain and back. While all models converge on the training set, not accounting for the above conversion leads to the divergence of the message decoders on the validation set for \cite{baluja2017hiding, zhu2018hidden}. }
\label{fig:single_msg_training_curve}
\end{figure*}

Notice, while both \cite{baluja2017hiding} and \cite{zhu2018hidden} yield low carrier errors, their direct application to speech data produced unintelligible messages with a low SNR. This is due to the fact that these models were not constrained to retain the same carrier content after the conversion to time-domain and back. Figure \ref{fig:single_msg_training_curve} depicts the training process of the proposed model and baselines. It can be seen that without any constraints, the baseline message decoders diverge. Lastly, Frequency Chop retains much of the message content after decoding, but creates a carrier with noticeable artifacts. This is due to the fact that the hidden message is audible as it resides in the carrier's high frequencies.

Moreover, we explored including adversarial loss terms between $\CF$ and $\tilde{\CF}$ to the optimization problem as suggested by~\cite{zhu2018hidden}. Similarly to the effect on images, when incorporating the adversarial loss, the carrier quality improved and contained less artifacts, however it comes with the cost of less accurate message reconstruction. 

Overall, the above results highlight the importance of modeling the time-frequency transformations in the context of steganographic models for the audio-domain.

\begin{table}[t]
\caption{Absolute Error (lower is better) and Signal to Noise Ratio (higher is better) for both carrier and message using single message embedding. Results are reported for both TIMIT and YOHO datasets.}
\label{tab:single_msg_res}
\centering
\resizebox{\columnwidth}{!}{%
\begin{tabular}{llcccc}
\toprule
& Model & Car. loss & Car. SNR & Msg. loss & Msg. SNR\\
\midrule
\multirow{5}{*}{\rotatebox{90}{TIMIT}} & Freq. Chop & 0.0770 & 0.22 & 0.046 & 6.85 \\ 
& Baluja \emph{et al.}~\cite{baluja2017hiding} & 0.0023 & 27.11 & 0.096 & 0.14 \\
& Zhu \emph{et al.}~\cite{zhu2018hidden} & 0.0027 & 32.70 & 0.078 & 0.71 \\
& Ours & \bf{0.0016} & 28.27 & \bf{0.035} & \bf{8.76} \\
& Ours + Adv. & 0.0022 & \bf{34.54} & 0.051 & 4.02\\
      \midrule 
\multirow{5}{*}{\rotatebox{90}{YOHO}} & Freq. Chop & 0.0550 & 0.24 & 0.038 & 7.08 \\
      & Baluja \emph{et al.} \cite{baluja2017hiding} & 0.0021 & 26.35 & 0.072 & 0.53 \\
      & Zhu \emph{et al.} \cite{zhu2018hidden} & 0.0047 & 27.99 & 0.066 & 1.05 \\
      & Ours & \bf{0.0016} & 27.86 & \bf{0.028} & \bf{8.16} \\
      & Ours + Adv. & 0.0016 & \bf{31.18} & 0.033 & 6.00 \\
\bottomrule
\end{tabular}
}
\end{table}

\paragraph*{Multiple messages.} Next, we further explore the capability of the proposed model for concealing several hidden messages. We analyzed the two settings described in Section~\ref{sec:model}, namely multiple decoders and single conditional decoder. Table~\ref{tab:multiple_msg_res} summarizes the results. The reported loss and SNR are averaged over the $k$ messages. Interestingly, both settings achieved comparable results for embedding 3 and 5 messages in a single carrier. 
An increase in the number of messages translates to higher loss values both for carrier and for messages. These results are to be expected as the model is forced to work at higher compression rates due to concealing and recovering more messages while keeping the carrier dimension the same. 

\begin{table}[t]
    \caption{AE and SNR for both carrier and message concealing 3 and 5 messages using either multiple decoders or one conditional decoder. Results are reported for both TIMIT and YOHO datasets.}    
\label{tab:multiple_msg_res}
\centering
\begin{tabular}{@{\extracolsep{4pt}}l|l|cc|cc}
\toprule
\multirow{6}{*}{\rotatebox{90}{TIMIT}} & & \multicolumn{2}{c|}{Carrier} & \multicolumn{2}{c}{Message}\\
\cline{2-6}
& model  & loss   & SNR   & loss   & SNR \\
\cline{2-6}
& multi-3 & 0.0042 & 25.13 & 0.0458 & 6.16 \\
& cond-3  & 0.0043 & 24.08 & 0.0463 & 6.08 \\
\cline{2-6}
& multi-5 & 0.0058 & 23.64 & 0.0550 & 4.42 \\
& cond-5  & 0.0063 & 22.70 & 0.0516 & 4.87 \\
\cline{1-6}
\multirow{4}{*}{\rotatebox{90}{YOHO}}
& multi-3 & 0.0042 & 23.80 & 0.0349 & 6.29 \\
& cond-3  & 0.0038 & 23.53 & 0.0344 & 6.49 \\
\cline{2-6}
& multi-5 & 0.0046 & 23.33 & 0.0428 & 4.17 \\
& cond-5  & 0.0051 & 22.30 & 0.0392 & 4.85 \\
\bottomrule
\end{tabular}
\label{tab:multiple_msg_res}
\end{table}

\section{Analysis}
\label{sec:analysis}
In this section we provide several evaluations regarding the quality of the embedded carrier, and the recovered message. We start with a subjective analysis of the resulted waveforms.

\subsection{Carrier ABX testing}
To validate that the difference between $\cf$ and $\tilde{\cf}$ is not detectable by humans, we performed ABX testing.
We present each human with two audio samples A and B. Each of these two samples is either the original carrier or the carrier embedded with a hidden message. These two samples are followed by a third sample X randomly selected to be either A or B. Next, the human must choose whether X is the same as A or B. We generated 50 (25 from TIMIT and 25 from YOHO) audio samples, for each audio sample we recorded 20 answers from Amazon Mechanical Turk (AMT), 1000 answers overall. Only 51.2\% (48.8\% for TIMIT and 53.6\% for YOHO) of the carriers embedded with hidden messages could be distinguished from the original ones by humans (the optimal ratio is 50\%). Therefore we conclude that the modifications made by the steganographic function are not detectable by the human ear.

\subsection{Message intelligibility}
A major metric in evaluating a speech steganography system is the intelligibility of the reconstructed messages. To quantify this measure we conducted an additional subjective experiment in AMT. We generated 40 samples from TIMIT dataset: 20 original messages and 20 messages reconstructed by our model. We used TIMIT for that task since it contains a reacher vocabulary set comparing to YOHO. We recorded 20 answers for each sample (800 answers overall). The participants were instructed to transcribe the presented samples, and the Word Error Rate (WER) and Character Error Rate (CER) were measured. While WER is a coarse measure, CER provides finer evaluation of transcription error. 
The CER/WER measured on original and reconstructed messages were 5.1\%/2.86\% and 5.15\%/2.78\% respectively. We therefore deduce that our system does not degrade the intelligibility of speech signal.

\subsection{Speaker recognition}
An advantage to concealing speech and not text is preservation of non-lexical content such as speaker identity. To evaluate that 
we conducted both human and automatic evaluations, adhering to the Speaker Verification Protocol \cite{heigold2016end}. Given 4 speech segments, the first three were uttered by a single speaker, the forth was uttered by either the same speaker or by a different one, the goal is to verify whether the speaker in the forth sample is the same as in the first three~\footnote{We use speakers of the same gender to make the task of speaker differentiation more challenging.}. For the human evaluation, we recorded 400 human answers, in 82\% of cases, listeners were able to distinguish whether the speaker in the forth sample matched the speaker in the first three. In the automatic evaluation setup, we used the automatic speaker verification system proposed by~\cite{heigold2016end}. The Equal Error Rate (EER) of the system is 18\% (82\% accuracy) on the generated messages, and 15\% EER (85\% accuracy) on the original messages. Hence, we deduce much of the speaker identity information is preserved in the generated messages. 

\subsection{Robustness to channel distortion}
Another critical evaluation is performance under noisy conditions. To explore that we applied different channel distortion and compression techniques on the reconstructed carrier $\tilde{\cf}$. In Table~\ref{tab:noise_msg_res} we describe message reconstruction results after distorting the carrier using: 16kHz to 8kHz down-sampling, MP3 compression (using different bit rates), 16-bit precision to 8-bit precision, Additive White Gaussian Noise (AWGN) and Speckle noise. 
Results suggest that our method is robust to carrier down-sampling, MP3 compression and noise addition. Contrarily, the model is sensitive to bit precision change, but this is to be expected as the message decoder relies on miniscule carrier modification in order to reconstruct the hidden message.

\begin{table}[t]
	\caption{Noise robustness results. We denote by $\sigma$ the norm of the added noise}
\label{tab:noise_msg_res}
\centering
\begin{tabular}{lcc}
\toprule
Noise & Msg. Loss & Msg. SNR \\
\midrule
Down-sampling to 8k     & 0.046 & 7.72\\
MP3 compression 128k  & 0.045 & 6.88 \\
MP3 compression 64k   & 0.062 & 5.34 \\
MP3 compression 32k   & 0.089  & 2.15 \\
AWGN, $\sigma=0.01$  & 0.077 & -12.52 \\
AWGN, $\sigma=0.001$ & 0.044  & 8.50   \\
Speckle, $\sigma=0.1$ & 0.035  & 8.26  \\
Speckle, $\sigma=0.01$ & 0.035  & 8.76 \\ 
Prec. reduction 8-bit & 0.160   & 0.25 \\
\bottomrule
\end{tabular}
\end{table}

\section{Related work}
\label{sec:rel_work}

A large variety of steganography methods have been proposed over the years, where most of them are applied to images \cite{morkel2005overview, kessler2004overview}.  Traditionally, steganographic functions exploited actual or perceptual redundancies in the carrier signal. The most common approach is to encode the secret message is in the least significant bits of individual signal samples~\cite{jayaram2011information}. Other methods include concealing the secret message in the {\em phase} of the frequency components of the carrier~\cite{dong2004data} or in the form of the parameters of a miniscule echo that is introduced into the carrier signal~\cite{bender1999method}.

Recently, neural networks have been widely used for steganography~\cite{baluja2017hiding, zhu2018hidden, hayes2017generating, qian2015deep, pibre2016deep, wu2018stegnet, tang2017automatic, el2008embedding, shi2017ssgan, cui2020multi, agarwal2020deep, zhang2019invisible}. 
The authors in \cite{baluja2017hiding} first suggested to train neural networks to hide an entire image within another image (similarly to Figure~\ref{fig:model}A). \cite{zhu2018hidden} extended the work of \cite{baluja2017hiding} while adding an adversarial loss term to the objective. \cite{hayes2017generating} suggested to use generative adversarial learning to generate stenographic images.  However, none of the above approaches explored speech data and were focused on hiding a single message only.

A closely related task is \textit{Watermarking}. Both approaches aim to encode a secret message into a data file. However, in steganography the goal is to perform secret communication while in watermarking the goal is verification and ownership protection. Several watermarking techniques use LSB encoding \cite{van1994digital, wolfgang1996watermark}. Recently, \cite{uchida2017embedding, adi2018turning} suggested to embed watermarks into neural networks parameters.

\section{Discussion and future work}
\label{sec:dis}

In this work we show that the recently proposed deep learning models for image steganography are less suitable for audio data. We show that in order to utilize such models, time-domain transformations must be addressed during training. Moreover, we extend the general deep-learning steganography approach to hide multiple messages. We evaluated our model under several noisy conditions and showed empirically that such modifications to carriers are indistinguishable by humans and the messages recovered by our model are highly intelligible. Finally, we demonstrated that voice speaker verification is a viable means of authentication for hidden speech messages. 

For future work we would like to explore the ability of such steganographic methods to evade detection by steganalysis algorithms, and incorporate such evasion capabilities as part of the training pipeline. 

{ 
\bibliographystyle{IEEEbib}
\bibliography{refs}
}

\end{document}